\documentclass{iau} 
\usepackage{graphicx}

\title[SB among AGB] 
{Spectroscopic binaries among AGB stars from HERMES/Mercator: the case of V Hya}

\author[Alain Jorissen et al.]   
{Alain Jorissen$^1$,
 Sophie Van Eck$^1$,
 Thibault Merle$^1$,
 \and Hans Van Winckel$^2$}

\affiliation{$^1$Institut d'Astronomie et d'Astrophysique, Universit\'e libre de Bruxelles, CP 226,\\
B-1050 Bruxelles, Belgium \\ email: {\tt ajorisse@ulb.ac.be} \\[\affilskip]
$^2$Institute of Astronomy, KU Leuven, Celestijnenlaan 200D, B-3001 Leuven, Belgium \\email: {\tt hans.vanwinckel@kuleuven.be}}

\pubyear{2018}
\volume{343}  
\jname{Why Galaxies Care About AGB Stars}
\editors{F. Kerschbaum, M. Groenewegen, \& H. Olofsson, eds.}
\begin{document}

\maketitle

\begin{abstract}
We report on our search for spectroscopic binaries among a sample of AGB stars. Observations were carried out in the framework of the monitoring of radial velocities of (candidate) binary stars performed at the Mercator 1.2m telescope, using the HERMES spectrograph. We found evidence for duplicity in UV Cam, TU Tau, BL Ori, VZ Per, T Dra, and V Hya. 
\end{abstract}


Several methods exist to find binaries: by detecting radial-velocity variations, eclipsing or ellipsoidal light curve, proper-motion anomalies, composite spectra, X-ray or UV flux excess, spiral arms in the circumstellar dust or molecular emission, binary signature in interferometric or lunar-occultation data (see the reviews by \cite[Jorissen 2004]{Jorissen2004}, and \cite[Jorissen \& Frankowski 2009]{JorissenFrankowski2009}). In the case of binaries involving mass-losing AGB stars, the duplicity sometime produces symbiotic activity. However, the fraction of AGB binaries exhibiting symbiotic activity is unknown. We therefore started a radial-velocity survey of non-symbiotic AGB stars. 
Very few spectroscopic binaries are known so far among non-symbiotic AGB stars. One such rare case is the M5III SRb variable (with $P_{\rm GCVS} = 44.3$~d) RR~UMi with $P_{\rm orb} = 749$~d \cite[(Batten \& Fletcher 1986)]{Batten1986}.
Spectroscopic binaries are difficult to find among  AGB stars because their  radial velocities bear signatures from regular envelope pulsation, or in less extreme (but more frequent) cases, atmospheric jitter. Both are the sources of annoying noise which may possibly mask the presence of orbital variations in the radial-velocity curve. Therefore, the detection of spectroscopic binaries is only possible if it triggers orbital velocity variations larger than this intrinsic velocity scatter (represented by the dashed line in Fig. 1, calibrated on a comparison sample of M giants; \cite[Jorissen \etal\ 2009]{Jorissenetal2009}).

Observations were carried out in the framework of the monitoring of radial velocities of (candidate) binary stars performed at the Mercator 1.2m telescope at La Palma \cite[(Gorlova \etal\ 2014)]{Gorlova2014}, 
using the HERMES high-resolution spectrograph \cite[(Raskin \etal\ 2011)]{Raskin2011}. 
The monitoring started in mid-2009, spans about 250 nights/yr, and involves AGB and carbon stars with suspicion of binarity (because of a composite spectrum, X-ray or UV flux, interferometric data, or lunar-occultation data;  \cite[Jorissen 2004]{Jorissen2004}, \cite[Sahai \etal\ 2008]{Sahai2008}).
We found 
evidence for duplicity in UV Cam, TU Tau, BL Ori, VZ Per, T Dra, and V Hya (large filled circles above the dashed line in Fig.~1). The situation for the carbon Mira  V~Hya is displayed in Fig.~2.
An orbital period of 8.5 yr is suspected by \cite[Sahai \etal\ (2016; also this volume)]{Sahai2016} from the dynamics of 'bullets' in jets, but this value is not confirmed by our monitoring which now spans just over 8.5~yr.  The orbital period looks instead similar to the long-term modulation observed in the AAVSO light curve
\cite[(Knapp \etal\ 1999)]{Knapp1999}, 
and is therefore likely caused by obscuration from a circumbinary disc, a characteristic feature of binary RV Tau stars of the b subtype \cite[(Manick \etal\ 2018)]{Manick2018}.

\begin{figure}[h]
\vspace*{-0.5 cm}
\begin{center}
\includegraphics[width=2in]{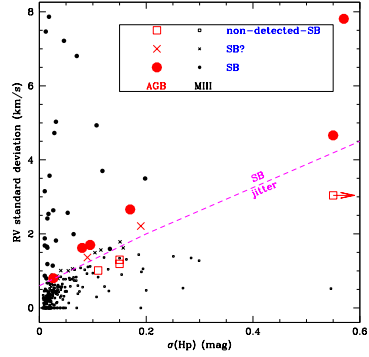}
   \label{fig1}
   \caption{The samples of candidate SB among AGB and carbon stars (large symbols), as compared to the reference sample of M giants (small symbols), from Jorissen et al. (2009), who identified SBs (small filled circles), possible SBs (small crosses), and non-SBs (small open squares). Spectroscopic binaries should thus be located above the dashed magenta line. $\sigma(Hp)$ is the standard deviation of the Hipparcos $Hp$ magnitude. V Hya is in the upper right corner.
}
\end{center}
\end{figure}

\begin{figure}[h]
\vspace*{-0.5 cm}
\begin{center}
\includegraphics[width=2.5in]{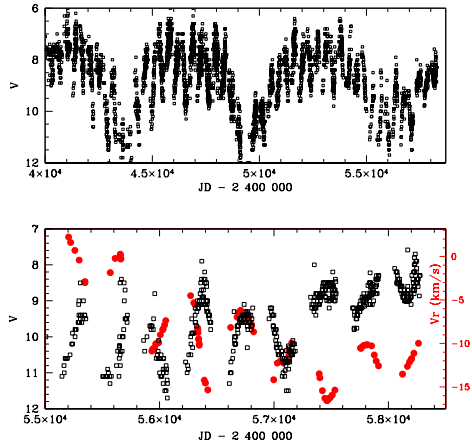} 
 \caption{Top panel: AAVSO light curve showing the long-term orbital modulation superimposed on the Mira pulsations. Bottom panel: Same as top with radial velocities added (filled circles). Note the clear phase shift between velocity and light curves.}
\vspace*{-0.5 cm}
   \label{fig2}
\end{center}
\end{figure}

\firstsection

\end{document}